\documentclass[twocolumn,showpacs,preprintnumbers,amsmath,amssymb]{revtex4}

\usepackage{graphicx}
\usepackage{dcolumn}
\usepackage{bm}
\usepackage{amssymb}

\newcommand\pp{{A'}}
\newcommand\g{{\gamma }}

\newcommand\vdecay{A' \rightarrow e^+  e^-}
\newcommand\xdecay{A' \rightarrow e^+  e^-}

\newcommand\pair{e^+ e^-}
\newcommand\ee{e^+e^-}

\newcommand\piga{\pi^0 \to \gamma A'}

\newcommand\pige{\pi^0 \to \gamma \ee}

\def\address{\@ifstar{\address@star}%
  {\@ifnextchar[{\address@optarg}{\address@noptarg}}}

\begin{document}

\author{S.N.~Gninenko}
\affiliation{Institute for Nuclear Research, Moscow 117312}


\title{Constraints on  dark photons from $\pi^0$ decays}

\date{\today}

\begin{abstract}
Several models of dark matter suggest the existence of hidden sectors consisting of $SU(3)_C \times SU(2)_L \times U(1)_Y$ singlet fields. 
The interaction between the ordinary  and hidden sectors could be  transmitted by 
new Abelian $U'(1)$ gauge  bosons $A'$ (dark or hidden photons)
 mixing with ordinary photons.     
If such $A'$'s have masses below the $\pi^0$ meson mass, they 
would be produced through  $\gamma - A'$ mixing  in the 
$\pi^0\to \gamma \gamma$ decays and be observed via decays 
$A' \to \ee $. Using bounds from the 
SINDRUM experiment at the Paul Scherrer Institute  that searched for an excess of $\ee$ pairs in $\pi^- p$ interactions at rest, the area excluding  the $\gamma - A'$ mixing    
$\epsilon \gtrsim 10^{-3}$ for  the  $A'$ mass  region 
$ 25 \lesssim M_{A'} \lesssim 120$ MeV is derived.  
\end{abstract}
\maketitle
The  origin of dark matter is still a  great puzzle in particle physics and
 cosmology. Several models  dealing with this problem suggest
the existence of `hidden' sectors consisting of $SU(3)_C \times SU(2)_L \times U(1)_Y$ singlet fields. These  sectors do not interact with our world directly 
and  couple to it by gravity. It is also possible that there exist new very-weak forces between the 
ordinary and dark worlds transmitted by  new Abelian $U'(1)$ gauge  bosons $A'$ (dark or hidden photons for short) mixing with our photons \cite{hop}, as discussed first by  Okun in his  model of 
 paraphotons \cite{okun}.  
In a class of recent interesting models the $\gamma-A'$  mixing strength may be large 
enough to be experimentally tested. 
This makes searches for $A'$'s very attractive; for a recent review see \cite{jr} and references therein.

It should be noted, that many models of physics beyond the  Standard Model (SM)  such as  GUTs~\cite{1}, superstring models~\cite{2} (see also Ref.\cite{khlop}), supersymmetric~\cite{3}, 
 and models including the fifth force \cite{carl} also predict an extra
U$^{'}$(1) factor and the corresponding new gauge $X$ boson.
The $X$'s could interact directly with quarks and/or leptons. 
If the $X$ mass  is below  the pion mass, the $X$ could be effectively
 searched for  in the decays $P\to \gamma X$, where $P = \pi^{0},\eta$, or 
$\eta^{\prime}$. This is due to the fact, that  the decay rate of  
$P\to \gamma~+~$ $\it any~new~particles~with~spin~0~or~\frac{1}{2}$ is proved
 to be negligibly small~\cite{di}. Hence, an observation of  these decay modes 
could unambiguously signal the discovery of a new spin-1 boson, in
contrast with other searches  for new light particles
in rare K, $\pi$ or $\mu$ decays~\cite{di,md,gkx2}.

The  allowed $\gamma - A'$ interaction is  given by the kinetic mixing \cite{okun,jr,holdom,foot1}  
\begin{equation}
 L_{int}= -\frac{1}{2}\epsilon F_{\mu\nu}A'^{\mu\nu} 
\label{mixing}
\end{equation}
where  $F^{\mu\nu}$, $A'^{\mu\nu}$ are the ordinary 
 and the  dark photon  fields, respectively, and $\epsilon$ is their mixing strength.  
In some recent dark matter models the dark photon could be massless; see, e.g. Refs.\cite{Cline:2012is,Cline:2012ei}. If the $A'$ has a mass, the  kinetic mixing of Eq.(\ref{mixing})
can be diagonalized resulting in a nondiagonal mass term and $\gamma - A'$ mixing. Hence, any $\gamma$-source could produce a kinematically allowed massive $A'$ boson  according to the appropriate mixings. 
Then, if the mass difference is small, ordinary photons may oscillate into dark photons-similarly to neutrino oscillations- or, if the mass difference is large, dark photons could  decay, e.g. into $\ee$ pairs. 

Experimental constaints on dark photons in the meV-keV mass range
 can be derived  from searches for the fifth force 
\cite{okun,c1,c2}, from experiments based  on the  photon 
regeneration technique \cite{phreg,bober,sik,rs,vanb}, and from astrophysical considerations \cite{seva1,seva2}. For example, the results of  experiments searching for solar axions \cite{cast1,cast2} 
can be used to set  limits on the $\g - \pp$ mixing in the keV part of the solar spectrum of dark photons
\cite{jr1,jr2,gr,st}. Stringent bounds on the low mass $A'$s
 could be obtained from astrophysical  considerations \cite{blin}-\cite{david}.  There are plans to  test the 
existence of sub-eV dark photons at new facilities,   such as, for example,  SHIPS \cite{ships} and IAXO \cite{igor}.

The $A'$'s with the masses in  the sub-GeV range, see e.g.
 \cite{bpr,rw,will}, can be searched for  through their $A'\to \ee$ decays in beam-dump experiments \cite{jdb,e137,brun,e141,e774,apex}, or in  particle 
decays \cite{bes,kloe,babar,mami}. Recently, stringent 
bounds on the mixing $\epsilon$  have been obtained from searches for  decay 
modes $\pi^0,\eta,\eta' \to \gamma A'(X)$, $A'(X)\to \ee$
 with existing data of neutrino experiments \cite{sngpi0,sngeta}. 
These limits are valid for  the relatively long-lived $A'$s with a 
mixing strength  in the range  $10^{-4}\lesssim \epsilon \lesssim 10^{-7}$. 
The goal of this note is to show  that new bounds on the decay $\pi^0 \to \gamma A'$ of neutral pions into a photon and a short-lived $A'$ followed by the rapid decay $A'\to \ee$ due to the relatively large 
$\gamma-A'$ mixing 
can be obtained  from the results of sensitive searches for an excess of single
isolated $\ee$ pairs from decays of the weakly interacting neutral boson $X$
  by the SINDRUM Collaboration at the Paul  Scherrer Institute (PSI, Switzerland) \cite{sindrum}.
  
The SINDRUM experiment- specifically 
designed to search for rare particle decays in the SINDRUM magnetic spectrometer- 
 was performed by using the  $\pi^- p $ interactions at rest as the source of 
$\pi^0$'s.  The $\pi^0$'s were  produced in the charge exchange reaction $\pi^- p \to \pi^0 n $ of 95 MeV/c $\pi^-$'s  stopped in a small liquid hydrogen target in the center of the 
SINDRUM magnetic spectrometer. The magnetic field was 0.33 T, resulting in a
transverse-momentum threshold of roughly 17 MeV/c for
particles reaching the scintillator hodoscope surrounding the target. The trigger
required an $\ee$ pair with an opening angle in the
plane perpendicular to the beam axis of at least 35$^o$; this
corresponds to a lower threshold in the invariant mass of 25 MeV/c \cite{sindrum}.
A  total  of 98 400  $\pige$ decays were observed. 
The signature of the $X\to \ee$ decay  would be seen as a peak
in the continuous $\ee$ invariant mass distribution. 

\begin{figure}
\includegraphics[width=0.5\textwidth]{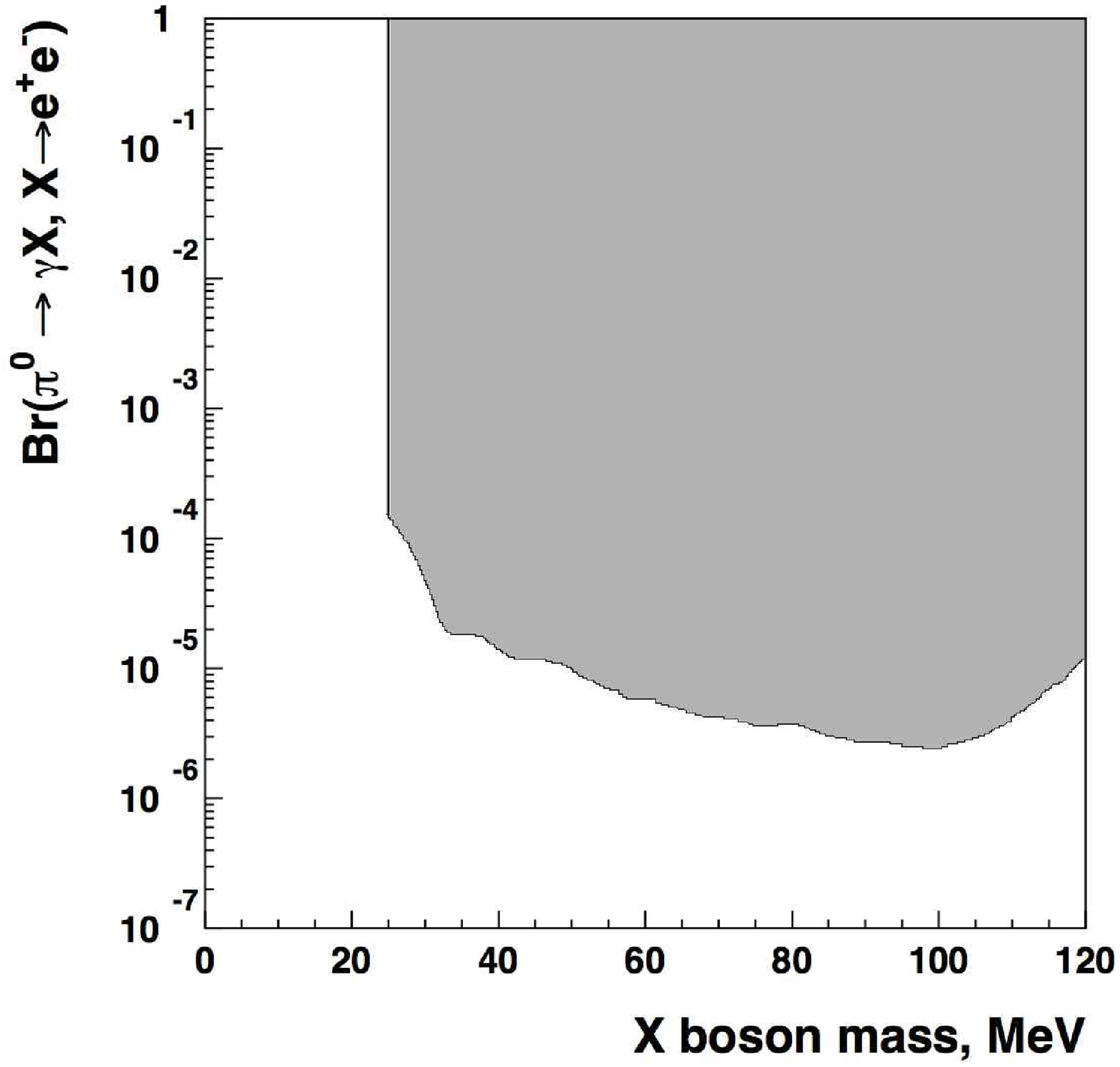}
\caption{ The 90 \% C.L. area (shaded)  in the $\bigl(M_{X}; Br(\pi^0\to \gamma X, X\to \ee)\bigr)$ plane excluded by the SINDRUM experiment (from Ref.\cite{sindrum}).}
\label{limit}
\end{figure}
No such peak events were found and upper limits on the branching ratio $Br(\pi^0\to \gamma X, X\to \ee)=\frac{\Gamma(\pi^0\to \gamma X, X\to \ee)}{\Gamma(\pi^0\to \gamma \gamma)}$  in the range  
$\simeq 10^{-6}-10^{-5}$ have been placed  for the  $X$-mass region $25 \lesssim M_X \lesssim  120$  MeV.
The corresponding  90\% C.L.  exclusion area  
in the $\bigl(M_{X}; Br(\pi^0\to \gamma X, X\to \ee)\bigr)$ plane 
is shown in Fig.\ref{limit}. The limits were obtained assuming  the $X$ lifetimes to be  in the range 
\begin{equation}
10^{-23} \lesssim \tau_{X} \lesssim 10^{-11} ~{\rm s}.
\label{lifetime}
\end{equation}
For lower values of $\tau_X$ in Eq.(\ref{lifetime}) the $\ee$ mass peak would be
smeared out beyond recognition; for larger values most $X$'s 
would decay outside the target region and thus the detector would not be triggered \cite{sindrum}.

If the $A'$ exists and is a short-lived particle, it would decay in the SINDRUM target and  be observed in the 
detector via the $\xdecay$ decay similar to the decays of 
$X$'s. The occurrence of $\xdecay$ decays  would appear as an excess
of $\pair$ pairs in the SINDRUM spectrometer above those expected from  
standard  decays of $\pi^0$ produced in $\pi^- p$ interactions.  
As  the final states of the  decays $\pi^0\to \gamma X, X\to  \ee$ and 
$\piga, \xdecay$ are identical, the  
results of the  searches for the former  can be used  
to constrain the latter for the same  $\ee$  invariant mass regions.
 
\begin{figure}
\includegraphics[width=0.5\textwidth]{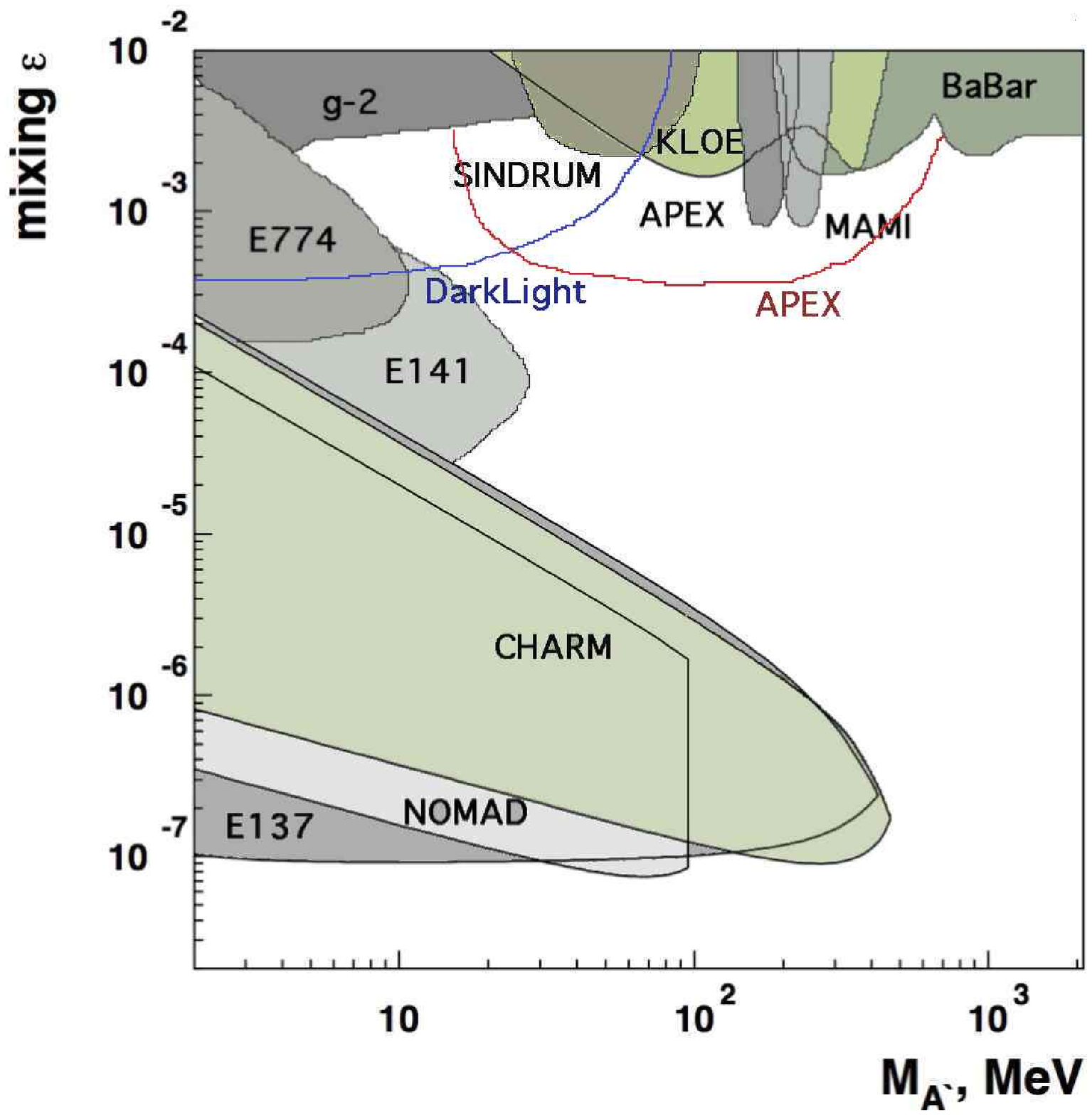}
\caption{ Exclusion region in the ($M_{A'}; \epsilon$) plane 
obtained in the present work from the results of the SINDRUM experiment 
\cite{sindrum}. Shown are areas excluded from the muon (g-2) considerations, by 
the results  of the electron beam-dump 
experiments E137 \cite{jdb,e137}, E141 \cite{e141}, E774 \cite{e774},
 the searches in APEX \cite{apex}, 
KLOE\cite{kloe}, BaBar\cite{babar}, and MAMI \cite{mami}, 
 and from the data of the neutrino 
experiments  NOMAD  \cite{sngpi0} and CHARM \cite{sngeta}. Expected sensitivities of the planned  APEX (full run)   and DarkLight experiments  are  also shown for comparison. For a review of all experiments, which 
intend   to probe a similar parameter space, see Ref.\cite{hif} and references therein.}
\label{plot}
\end{figure}
For a given number $N_{\pi^0}$ of $\pi^0$'s produced in the target  the expected number of $\xdecay$ (or $X\to \ee$)  decays 
occuring within the fiducial volume  of the 
SINDRUM detector  is given by 
\begin{eqnarray}
N_{\xdecay}(M_{A'}) = \int f\Bigl[1-{\rm exp}\Bigl(-\frac{r M_{A'}}{P\tau_{A'}}\Bigr)\Bigr]\zeta A dr d\Omega \nonumber \\
=N_{\pi^0}Br(\piga) Br(A' \to\ee)\zeta A~~
\label{nev}
\end{eqnarray}
where $M_{A'},~ P, ~f ,~r, ~\tau_{A'}$ are the $A'$ mass, momentum, flux, the distance between the $A'$ decay vertex and the target,  and the lifetime  
at rest, respectively and  $\zeta$ and $A$  are the $\pair$ pair reconstruction efficiency
and the acceptance  of the SINDRUM spectrometer, respectively \cite{sindrum}. 
Here it is assumed that  the $A'$ is a short-lived particle with 
$\frac{r M_{A'}}{P\tau_{A'}} \gg 1 $ for $r$ values larger than the effective size of the target, in accordance with Eq.(\ref{lifetime}). 
Taking  Eq.(\ref{nev}) into account  and using the relation
 $ N_{\xdecay}(M_{A'}) < N_{\ee}^{90\%}(M_{A'}) $, where $N_{\ee}^{90\%}(M_{A'})$  is the 
90\% C.L. upper limit for the  number of signal events from the 
decays of the $A'$ with a  given mass $M_{A'}$,  results in  the $90\%$ C.L. exclusion area in the 
($M_{A'};Br(\piga, A' \to\ee) $) plane obtained by  the SINDRUM experiment and shown  in Fig.\ref{limit}.
  The upper limit  $N_{\ee}^{90\%}$ as a function of $M_{A'}$ was obtained  from 
the fit of the measured  $\ee$ mass distribution  in the vicinity of each selected value of  $M_{A'}$,   
to a sum of the signal peak from the $A'\to \ee$ decays and a flat background distribution.

The obtained results can be used to impose bounds on the $\gamma-A'$ mixing strength 
as a function of the  dark photon mass. 
For $A'$ masses smaller than  the mass $M_{\pi^0}$ of 
the $\pi^0$ meson, the branching fraction of  the  decay 
$\pi^0 \to \gamma A'$ is   given by \cite{bpr}: 
\begin{equation}
Br(\pi^0 \to \gamma A') = 2\epsilon^2 Br(\pi^0 \to \gamma \gamma) \Bigl( 1- \frac{M_{A'}^2}{M_{\pi^0}^2}\Bigr)^3.
\label{br}
\end{equation}
Assuming that   the dominant $A'$-decay is into a $\ee$ pair,
the corresponding decay rate is given by:
\begin{equation}
\Gamma (\vdecay) = \frac{\alpha}{3} \epsilon^2 M_{A'} \sqrt{1-\frac{4m_e^2}{M_{A'}^2}} \Bigl( 1+ \frac{2m_e^2}{M_{A'}^2}\Bigr)
\label{rate}
\end{equation}

Taking into account Eq.(\ref{br}),   
one can  determine the $90\%$ C.L. exclusion area in the 
($M_{A'}; \epsilon $) plane from the results of the SINDRUM experiment. This 
area is shown in Fig. \ref{plot}, together with regions excluded by the results of  the  electron beam-dump experiments E137, E141, E774  \cite{jdb,e137,e141,e774}, by recent measurements from APEX \cite{apex}, KLOE \cite{kloe}, BaBar \cite{babar}, and MAMI \cite{mami}, and from the data of the neutrino experiments NOMAD \cite{sngpi0} and CHARM \cite{sngeta}.  For a recent, 
more detailed review of existing and planned limits, see Refs. \cite{hif,sarah1,sarah2}. 
The shape of the exclusion contour  from the  SINDRUM experiment corresponding 
to the $A'$ masses  $M_{A'} \gtrsim$ 100 MeV is defined mainly by the phase-space factor in Eq.(\ref{br}). The  $A'$ lifetime values 
 calculated by using Eq.(\ref{rate}) for the mass range  
$25 \lesssim M_X \lesssim 120$ MeV are found to be within the allowed range of 
Eq.(\ref{lifetime}). Note, that since the $A'$ is a short-lived particle, the sensitivity 
of the search is $\propto \epsilon^2$, differently from the case of a long-lived $A'$, where 
the number of signal events is $\propto \epsilon^4$; see, e.g. Refs.\cite{sngpi0,sngeta}.

In summary, using results from the SINDRUM experiments on the search for weakly interacting $X$ bosons produced in $\pi^- p$ interactions at rest and decaying 
into $\ee$ pairs, new bounds on a hidden-sector gauge  $A'$ boson 
produced in the decay $\pi^0 \to \gamma A'$  were derived.
The obtained exclusion area  covers the $A'$ mass region $ 25 \lesssim M_{A'}\lesssim 120$ MeV  and   the $\gamma - A'$ mixing strength  $  \epsilon \gtrsim 10^{-3}  $. 
\vskip0.2cm

 The help of D. Sillou in calculations is greatly appreciated.

\end{document}